%% file: aps.tex
\documentclass[%
 reprint,
 amsmath,amssymb,
 aps,
]{revtex4-2}

\usepackage{graphicx}
\usepackage{dcolumn}
\usepackage{bm}

\usepackage{xcolor}

\newcommand{\dd}{\mathrm{d}}
\newcommand{\dv}[2]{\frac{\dd #1}{\dd #2}}

\usepackage{nicefrac}   

\usepackage[ruled,vlined]{algorithm2e}

\begin{document}

\preprint{APS/123-QED}

\title{Allostery Beyond Amplification: Temporal Regulation of Signaling Information}

\author{Pedro Pessoa$^{1,2}$}
\author{Steve Press\'e$^{1,2,3}$}
\author{S.~Banu Ozkan$^{1,2}$}

\affiliation{$^{1}$ Center for Biological Physics, Arizona State University, Tempe, AZ, USA}
\affiliation{$^{2}$ Department of Physics, Arizona State University, Tempe, AZ, USA}
\affiliation{$^{3}$ School of Molecular Sciences, Arizona State University, Tempe, AZ, USA}



\begin{abstract}
Allostery is a fundamental mechanism of protein regulation and is commonly interpreted as modulating enzymatic activity or product abundance. Here we show that this view is incomplete. Using a stochastic model of allosteric regulation combined with an information-theoretic analysis, we quantify the mutual information between an enzyme’s regulatory state and the states of downstream signaling components. Beyond controlling steady-state production levels, allostery also regulates the timing and duration over which information is transmitted. By tuning the temporal operating regime of signaling pathways, allosteric regulation enables distinct dynamical outcomes from identical molecular components, providing a physical mechanism for temporal information flow, signaling specificity, and coordination without changes in metabolic pathways.

\end{abstract}

\maketitle


\input{text}

\bibliography{refs}

\newpage 
\onecolumngrid
\appendix

\include{appendixs}

\end{document}

%% file: text.tex
\section{Introduction}

Allostery is a fundamental mechanism of remote regulation in proteins, whereby ligand binding at one site modulates activity at a distant functional site \cite{Phillips20}. Reflecting on its discovery, Monod famously referred to allostery as ``the second secret of life” \cite{Monod71}, underscoring its deep biological significance. While decades of work have focused on how allosteric ligands reshape conformational free-energy landscapes \cite{Tsai14,Rouviere23,Nussinov25,Vossel25}, this thermodynamic perspective alone does not explain a striking empirical observation: Nearly every biochemical signaling pathway is composed predominantly of allosterically regulated proteins \cite{Gunasekaran04,Mathy23, Hamilton24}.

This ubiquity raises a fundamental and largely unaddressed question: Why has evolution so consistently selected allostery as the dominant regulatory strategy in signaling networks? Beyond simply modulating the steady-state level of product formation, allosteric proteins function as circuit components that transmit information between pathway elements by controlling the timing, duration, and ordering of signaling events. In signaling cascades, proteins must not only respond correctly but do so on appropriate timescales and in the correct temporal order. We hypothesize that allostery provides a physical mechanism for this control by coupling ligand binding to conformational changes that propagate signals with specificity and temporal precision.

G-protein-coupled receptors (GPCRs) illustrate this principle in fast signaling pathways \cite{Culhane22}. Ligand binding induces rapid conformational transitions that activate downstream effectors within milliseconds, enabling timely cellular responses. More generally, allosteric regulation enhances sensitivity to weak stimuli and allows precise control over ``when" downstream components are engaged and ``for how long". In this sense, allostery is inherently kinetic: it governs the timing, duration, and sequencing of signal transmission rather than merely shifting equilibrium activity levels.

The same principle extends beyond fast signaling to biological processes that require long-term temporal coordination. A striking example is the KaiABC circadian clock in cyanobacteria \cite{Kbler24}, where a network of allosteric protein–protein interactions generates robust, ~24-hour oscillations in phosphorylation state in the absence of transcriptional feedback. In this system, allosteric transitions in KaiC, modulated by KaiA and KaiB, encode timing information through ordered conformational cycles rather than changes in equilibrium activity. The KaiABC oscillator demonstrates that allostery can regulate information flow across vastly different timescales, from milliseconds to hours, highlighting its role as a universal mechanism for temporal control in biological networks.

Despite this central role, there is currently no unifying physical framework that explains how allostery enables information transmission along signaling pathways in time. Existing kinetic models are typically tailored to specific systems and focus on steady-state behavior or individual reaction steps, offering limited insight into how temporal coupling emerges between upstream and downstream components. As a result, the physical principles by which allosteric regulation coordinates signaling dynamics across multiple proteins remain poorly understood.

Here, we propose that the primary evolutionary advantage of allostery lies in its ability to regulate temporal information flow in biochemical networks. To test this hypothesis, we develop an information-theoretic framework based on a generalizable chemical master equation (CME) formalism \cite{VanKampen,Presse23,Pessoa24}. By computing the mutual information between an allosterically modulated upstream enzyme and a downstream signaling protein activated by its product, we show that allostery controls not only activity but also the timing, magnitude, and duration of coupling between pathway components. 

Through this framework, we demonstrate how the time-dependent dynamics of substrate concentration modulate the duration of coupling between signaling proteins, enabling the coordinated orchestration of cellular processes across diverse environmental contexts. By regulating substrate availability, cells can tune the temporal window over which upstream and downstream proteins remain coupled, ensuring that signaling dynamics align with specific temporal and spatial requirements. 
Concretely, we compute the mutual information \cite{Shannon48,Cover91} to quantify how much uncertainty about the enzyme’s regulatory configuration  --  capturing both conformation (allosteric vs. non-allosteric) and occupancy (bound vs. unbound) -- is reduced by observing the state of the downstream protein. 
Mutual information provides a natural measure of effective information flow through the pathway \cite{Presse13,Pessoa21}, capturing how allosteric regulation modulates the fidelity with which transient enzyme states are encoded into downstream molecular dynamics. 
This dynamic control provides a mechanistic basis for how signaling pathways achieve adaptability and robustness, highlighting the central role of allostery in regulating cellular behavior through temporal information flow \cite{Tkacik11,Mehta15,Modi20,Vossel25}.

\section{Methods}\label{sec:methods}
We model allosteric regulation using a minimal enzymatic reaction network formulated within a chemical master equation (CME) framework \cite{Modi20,Koshland66}. 
The system consists of an enzyme $A$ that catalyzes the conversion of a substrate $S$ into a product $P$, and a downstream protein $B$ that binds the product to form the complex $BP$. The enzyme $A$ can exist in two conformational states: a baseline state $A$ and an allosterically modified state $A^\ast$.

Both conformational states can bind substrate to form the complexes $AS$ and $A^\ast S$, but they may differ in substrate affinity and catalytic turnover. Using standard definition of the allosteric state \cite{Tsai14,Phillips20} the baseline conformation $A$ is more probable in the unbound state, whereas substrate binding stabilizes the allosterically modified conformation $A^\ast$. This coupling between binding and conformation provides a minimal kinetic representation of allosteric regulation.

\begin{figure}
    \centering
    \includegraphics[width=\linewidth]{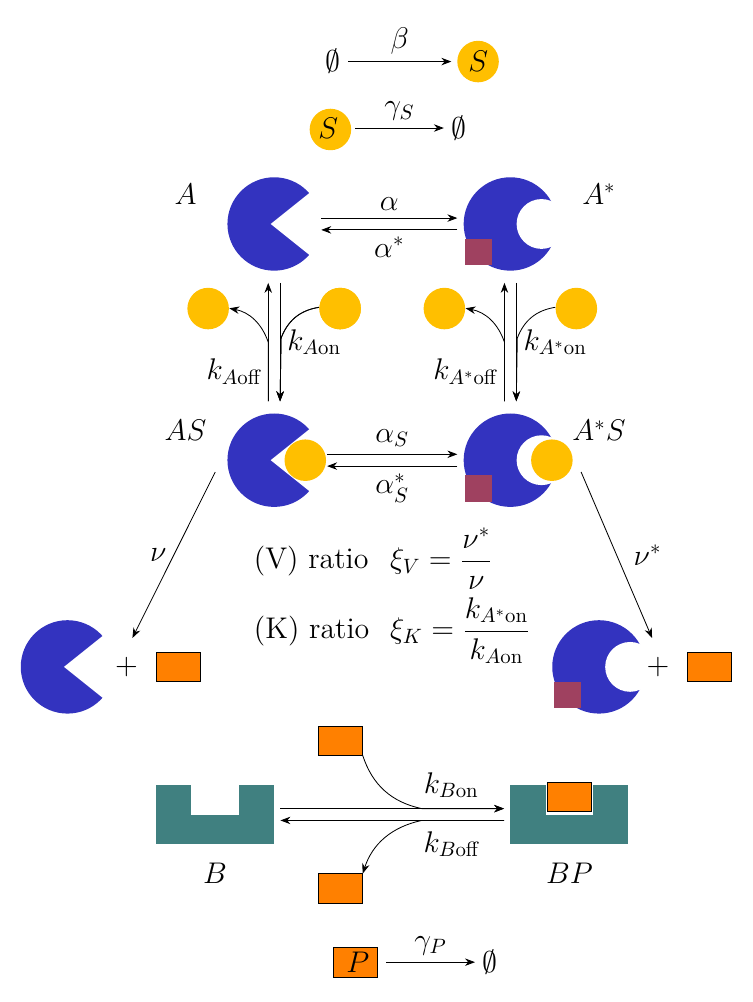}
    \vspace{-.5cm}
    \caption{\textbf{Diagram of the allosteric reaction network.}
    The substrate $S$ is produced at rate $\beta$ and degraded at rate $\gamma_S$, and interacts with the upstream (sender) enzyme $A$. 
    Enzyme $A$ occupies four internal states: unbound baseline ($A$), unbound allosterically modified ($A^\ast$), substrate-bound baseline ($AS$), and substrate-bound allosterically modified ($A^\ast S$).  
    The state $A^\ast$ represents an allosterically modified conformation stabilized by binding of an allosteric ligand (shown in brown). Allosteric switching between $A$ and $A^\ast$ modulates both substrate affinity and catalytic turnover. K-type allostery is quantified by the ratio $\xi_K = k_{A^\ast \mathrm{on}}/k_{A\mathrm{on}}$, which captures changes in substrate association rates, while V-type allostery is quantified by the ratio $\xi_V = \nu^\ast/\nu$, which captures changes in product generation rates. 
    The product $P$ is either degraded at rate $\gamma_P$ or sequestered by the downstream (receiver) enzyme $B$, which transitions between the unbound state $B$ and the product-bound state $BP$.} \vspace{-.6cm}
    \label{fig:reaction_model}
\end{figure}

To capture downstream signal processing, the model includes sequestration of the product $P$ by protein $B$. This downstream interaction allows us to quantify how product consumption influences the effective signaling output and temporal coupling between upstream and downstream components. A schematic of all molecular species and reactions is shown in Fig.~\ref{fig:reaction_model}.

Among the kinetic variables defined in Fig.~\ref{fig:reaction_model}, we quantify allosteric regulation using two dimensionless ratios. 
K-type allostery modifies substrate affinity, while V-type allostery alters catalytic turnover rate. 
Accordingly, we define
\begin{equation}
\xi_{K} \equiv \frac{k_{\mathrm{on}}^{A^\ast}}{k_{\mathrm{on}}^{A}}
\qquad\text{and}\qquad
\xi_{V} \equiv \frac{\nu^\ast}{\nu},
\end{equation}
where the K-type allosteric ratio, $\xi_{K}$, quantifies the change in substrate association rate between the allosteric and baseline states, and the V-type allosteric ratio, $\xi_{V}$, quantifies the corresponding change in catalytic rate. 
Values $\xi>1$ correspond to cooperative allostery, whereas $\xi<1$ indicate inhibition.

Having defined the reaction network and the two forms of allosteric regulation encoded by $\xi_K$ and $\xi_V$, we next examine how variations in these allosteric ratios shape the flow of information through the system within a fully stochastic framework. 
To do this, we track the joint stochastic time evolution of molecular copy numbers together with the internal conformational states of the enzymes.

The state of enzyme $A$ is represented by the discrete variable $\sigma_A \in \{A, A^\ast, AS, A^\ast S\}$, indicating whether the enzyme is unbound or substrate-bound, and whether it occupies the baseline or allosteric conformation. 
Similarly, the state of downstream protein $B$ is represented by $\sigma_B \in \{B, BP\}$, distinguishing whether it is free or product-bound. 
We consider a single copy $A$ and $B$. Such that the stochasticity arises from state switching and molecular turnover rather than from fluctuations in enzyme copy number.
Together with the copy numbers of $S$ and $P$, these variables define the full system configuration.

The time evolution of the joint probability distribution $p(\sigma_A, \sigma_B, S, P)$ reflects the coupled dynamics of binding, catalysis, allosteric switching, and product sequestration. 
These dynamics are governed by the chemical master equation (CME), which enumerates all possible state transitions according to the reaction rates in Fig.~\ref{fig:reaction_model}. 
We assume a well-mixed environment with fixed system volume and constant enzyme copy numbers, so that the dynamics is Markovian and fully described by the CME. 
The explicit CME corresponding to this network is derived in Supplementary Information section~\ref{SIsec:cme}.

In the results presented below, we vary the substrate generation rate $\beta$ together with the allosteric ratios $\xi_K$ and $\xi_V$ to examine how distinct modes of allosteric regulation propagate through the signaling network. 

All kinetic parameters are expressed in units of the substrate degradation rate $\gamma_S$, which sets the fundamental timescale of the system. We therefore fix $\gamma_S = 1$, so that all remaining rates should be interpreted as relative to substrate turnover. Our focus is on qualitative trends in information transmission rather than fine-tuned realism. Accordingly, all kinetic processes are chosen to be comparable to $\gamma_S$.   

Throughout the Results, the remaining kinetic parameters are fixed as
$k_{A\text{on}} = 1$, $k_{A\text{off}} = 1$, $k_{A^\ast\text{on}} = \xi_K$, $k_{A^\ast\text{off}} = 1$, $\alpha = 1$, $\alpha^\ast = 2$, $\alpha_S = 4$, $\alpha_S^\ast = 1$, $k_{B\text{on}} = 1$, $k_{B\text{off}} = 1$, and $\gamma_P = 1$, while the substrate production rate $\beta$ is varied as indicated in each figure. Unless otherwise noted in the figure captions, we set $\nu = 1$ and $\nu^\ast = \xi_V$. 
The specific choices $\alpha^\ast = 2$ and $\alpha_S = 4$ impose motivated by the previously mentioned stability order: in the absence of substrate, the non-allosteric conformation $A$ is more stable than $A^\ast$, while the allosteric confirmation $A^\ast S$ more stable than $AS$. This minimal energetic asymmetry captures the hallmark thermodynamic signature 
\cite{Tsai14,Phillips20}.

To solve the CME and obtain the time evolution and stationary joint distribution of all species, we employ numerical methods based on the theory of Markov jump processes. 
Details of the algorithms are provided in Supplementary Information section~\ref{SIsec:solution}. 
These methods were previously identified as high-performing in a recent benchmarking study \cite{Pessoa24}. The full code leading to all figures in the present article are available in our GitHub repository \cite{github}.

With the time evolution of the joint probability distribution established, we quantify information flow between enzymes by measuring the statistical dependence between their internal states $\sigma_A$ and $\sigma_B$. 
We compute the mutual information
\begin{equation}\label{mi_def}
\text{MI}_{AB}=\sum_{\sigma_A} \sum_{\sigma_B}
p(\sigma_A,\sigma_B) \log\!\left[ \frac{ p(\sigma_A,\sigma_B) }{ p(\sigma_A)\,p(\sigma_B) } \right],
\end{equation}
where, unless stated otherwise, the joint distribution $p(\sigma_A,\sigma_B)$ is evaluated at steady state and obtained by marginalizing the full distribution $p(\sigma_A, \sigma_B, S, P)$. Throughout, $\log$ denotes the natural logarithm.
Small values of $\text{MI}_{AB}$ indicate weak statistical dependence between enzymes, whereas larger values indicate stronger coupling and thus more effective communication. 
Alongside $\text{MI}_{AB}$, we also evaluate the expected copy numbers $\langle S \rangle$ and $\langle P \rangle$ to track how substrate usage and product output depend on allosteric regulation. 
Details on how each metric is calculated in the CME scheme previously described can be found in Supplementary Information section~\ref{SIsec:solution}.

\section{Results}\label{sec:results}
We begin by examining how K- and V-type allosteric regulation shape the steady-state coupling between enzymes $A$ and $B$. 
For each set of allosteric ratios, we solve the CME to obtain the steady-state distribution, meaning the stationary distribution after all components reach chemical equilibrium, and evaluate $\text{MI}_{AB}$ across a range of substrate input rates $\beta$.

\begin{figure}
    \centering
    \includegraphics[width=.95\linewidth]{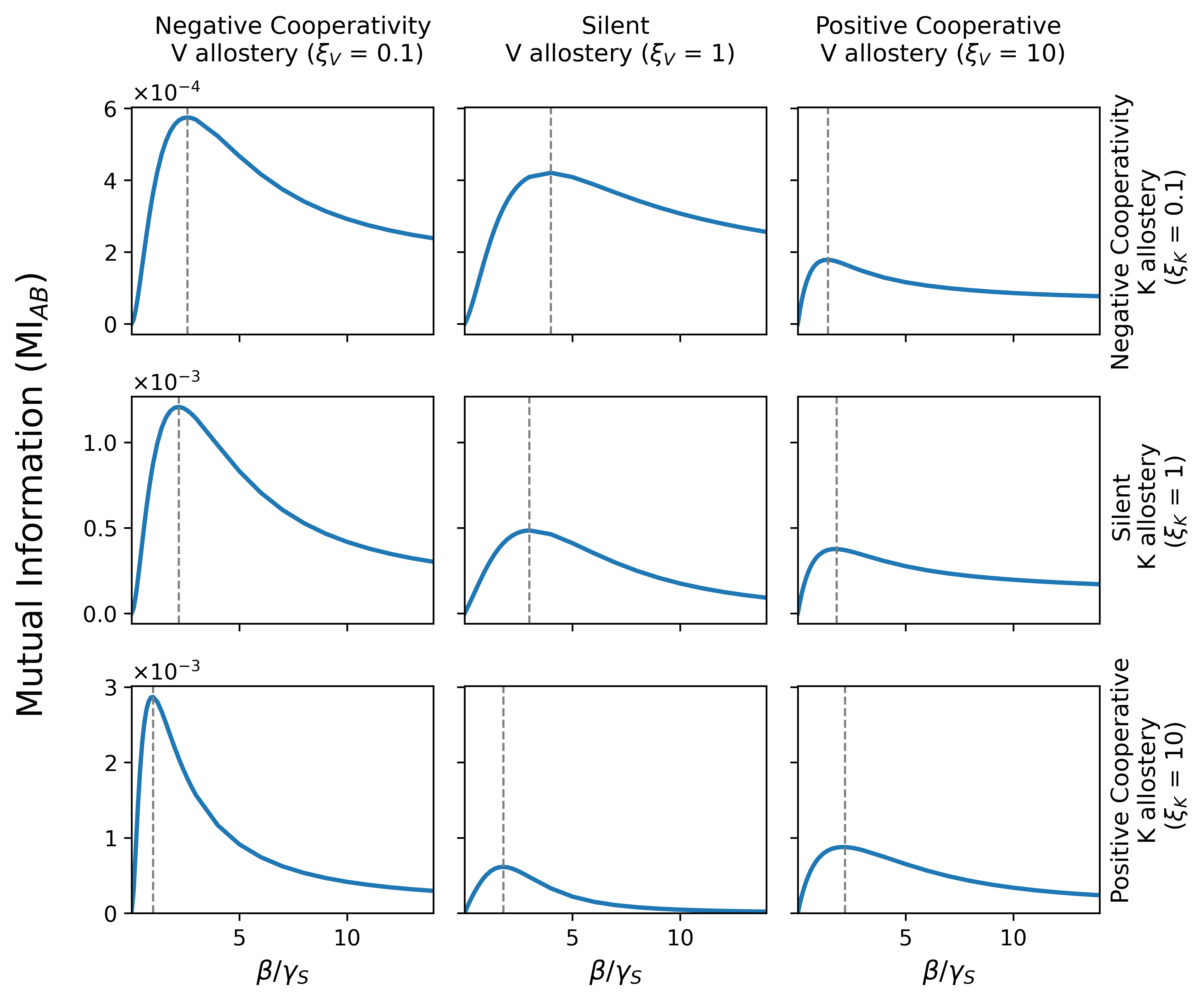} \vspace{-.5cm}
    \caption{
    \textbf{Calibrating the V- and K-type allosteric ratios allows for fine-tuning of the communication between the enzyme $A$ and the downstream protein $B$.} 
    The steady-state mutual information $\mathrm{MI}_{AB}$ as a function of the normalized substrate production rate $\beta/\gamma_S$ is shown for different combinations of K-type ($\xi_K$) and V-type ($\xi_V$) allostery. For all allosteric regimes, information transmission is maximized at an intermediate substrate flux, reflecting a balance between insufficient coupling at low flux and downstream saturation at high flux. The substrate flux that maximizes $\mathrm{MI}_{AB}$ (vertical dashed lines) depends sensitively on the allosteric parameters, demonstrating that K- and V-type allostery shift both the magnitude of transmitted information and the system’s optimal operating point. }\vspace{-.6cm}
    \label{fig:MI_steady}
\end{figure}
As shown in Fig.~2, $\mathrm{MI}_{AB}$ exhibits a robust non-monotonic dependence on the normalized substrate flux $\beta/\gamma_S$ across all allosteric conditions. At low substrate concentration,  mutual information  is low because both $A$ and $B$ are rarely engaged, leading to sparse and uncorrelated activity. At high substrate supply, communication again diminishes due to saturation of the upstream enzyme, which becomes persistently occupied and insensitive to time arrival of the product and substrate availability. Consequently, maximum mutual information occurs at an intermediate substrate flux, where product generation is sufficiently frequent to couple $A$ and $B$ without saturating the network.

Importantly, the position of this optimum depends on the form of allosteric control. Modulating $\xi_K$ or $\xi_V$ shifts the substrate level at which the information tranmission between the enzymes A and protein B is maximized. This shows   that allostery primarily tunes the operating point of the coupled enzymatic process rather than simply amplifying or suppressing signaling.

Having shown that information transmission between the upstream enzyme $A$ and the downstream protein $B$ depends nonmonotonically on the substrate production rate $\beta$ for different forms of allosteric regulation (Fig.~\ref{fig:MI_steady}), we now address the  next complementary question: how allosteric regulation itself reshapes steady-state behavior when the substrate input rate is held fixed. Rather than varying substrate supply, here we treat $\beta$ as a control parameter and systematically tune the allosteric properties of the enzyme.

Specifically, we fix one allosteric ratio at its neutral value ($\xi = 1$) and vary the other by several orders of magnitude, spanning regimes of strong inhibition to strong activation. For each condition, the system is evolved to steady state. This protocol allows us to isolate how $K$-type and $V$-type allostery independently influence information transmission, substrate availability, and product accumulation under identical substrate input conditions.

\begin{figure}
    \centering
    \includegraphics[width=.95\linewidth]{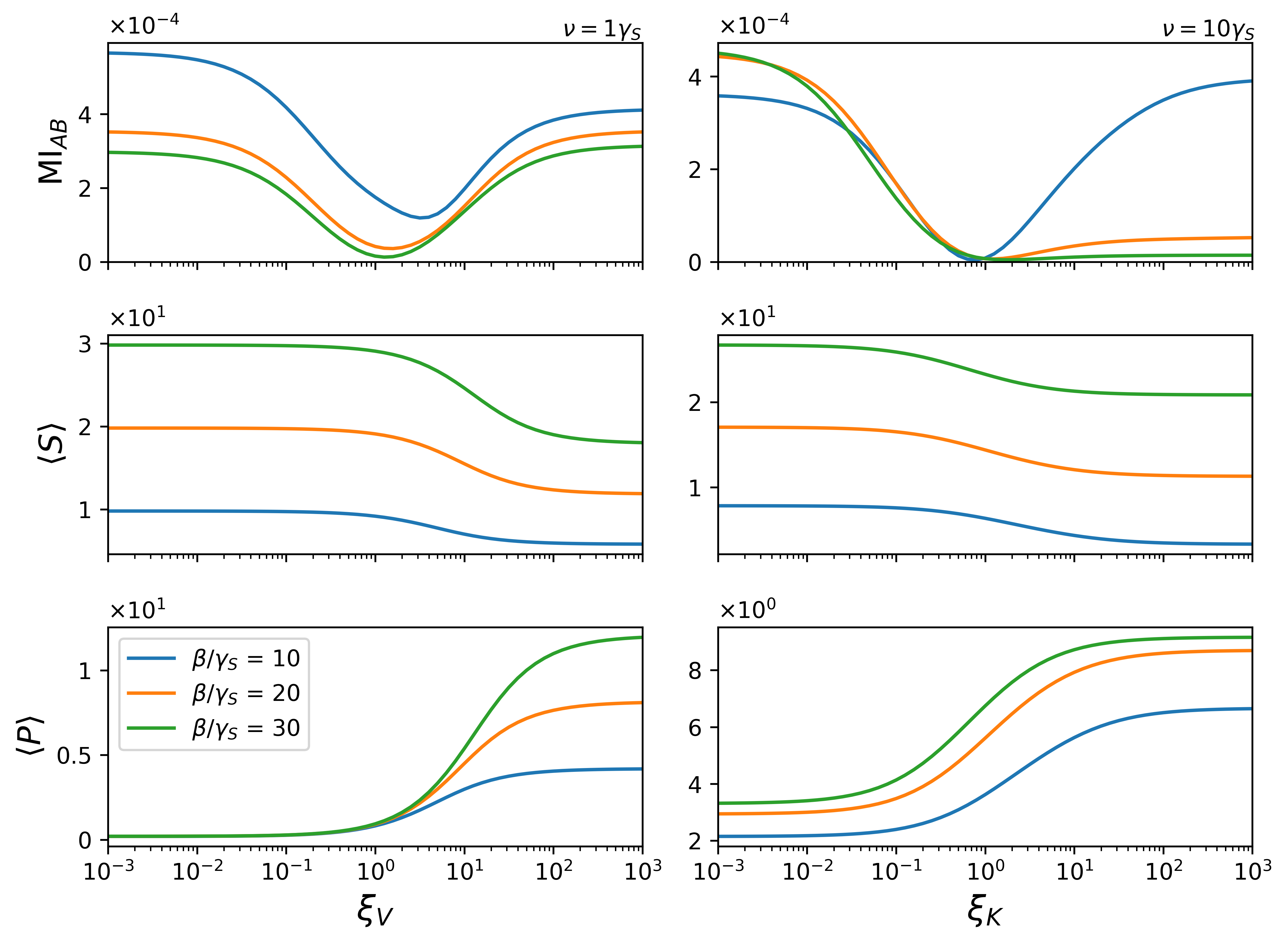}\vspace{-.5cm}
    \caption{
    \textbf{Allostery is not simply about making more product. }  
    Steady-state metrics across a range of relative production rates, $\beta/\gamma_S$.   
    The metrics shown, from top to bottom, are the mutual information between the states of $A$ and $B$ ($\text{MI}_{AB}$), the expected amount of substrate $\langle S \rangle$, and the expected amount of product $\langle P \rangle$ in the environment, all calculated at steady state.     
    For a silent K-type allosteric, $\xi_K=1$, across all values of $\beta/\gamma_S$, we observe that mutual information increases at both low (negative cooperativity) and high (positive cooperativity) V-type allosteric ratio, $\xi_V$. This increase in mutual information at high V-type allosteric ratio can be attributed to the rapid production rate, where $A$ and $B$ become sequentially active, and this is reflected in the variations of $\langle S \rangle$ and $\langle P \rangle$.     
    Conversely, the increase in mutual information at low allosteric (in both cases) rates indicates that even when allostery represents inhibition and, therefore, there is low production of $P$, communication between $A$ and $B$ is stronger.} \vspace{-.6cm}
    \label{fig:main_steady}
\end{figure}
Fig.~\ref{fig:main_steady} presents  the resulting steady-state behavior. For each parameter set, we report the mutual information between the internal states of enzymes $A$ and $B$, $\text{MI}_{AB}$, together with the mean substrate and product copy numbers, $\langle S \rangle$ and $\langle P \rangle$. Changes in $\text{MI}_{AB}$ are not simply correlated with changes in product abundance. Instead, mutual information is enhanced in two distinct dynamical regimes: one corresponding to strong catalytic activation and the other to strong inhibition.

In the high-$\xi_V$ regime, rapid catalytic turnover leads to sequential activation of $A$ and $B$, generating coordinated fluctuations that enhance statistical coupling between their states. By contrast, in the strongly inhibitory regime, slow catalysis prevents immediate saturation and maintains fluctuating product availability, which also strengthens coupling despite reduced mean product levels. In both limits, enhanced communication arises from dynamical coordination rather than from increased product output.

Taken together, these results demonstrate that allostery does not act merely as a gain control on product formation. Instead, allosteric regulation tunes the dynamical regime in which signaling operates by controlling when fluctuations in upstream activity are most effectively transmitted downstream. In our model, modest changes in allosteric parameters shift the substrate input rate at which mutual information is maximized, effectively retuning the temporal operating point of the same signaling architecture.

This mechanism is consistent with biological observations that signaling pathways governing fundamental processes such as cell-cycle progression and developmental timing are highly conserved across species ranging from amphibians to mammals \cite{GarciaOjalvo23,Rayon20}, yet exhibit markedly different temporal dynamics and response timescales \cite{Rayon20,Campbell24,Iwata24}. Importantly, these differences often arise without changes to core biochemical pathway topology, suggesting that evolutionary adaptation frequently proceeds through quantitative modulation of regulatory parameters rather than wholesale rewiring of signaling networks.

Our results provide a physical basis for this form of evolutionary tuning: by adjusting allosteric coupling strengths, conserved biochemical pathways can be retuned to operate optimally under different substrate input regimes, thereby supporting diverse physiological timescales and environmental demands. Such tuning of dynamical regimes offers a parsimonious route to functional diversification while preserving core pathway structure, consistent with recent evidence that signaling specificity and timing can be reshaped through parameter modulation alone \cite{RazoMejia18}.

Having established that allosteric regulation controls steady-state information transmission by tuning the operating point of the system, we next asked how these same mechanisms regulate dynamical coupling in response to time-dependent inputs. In many cellular contexts, substrate availability is not constant but fluctuates due to environmental changes, metabolic cycles, or upstream regulatory events. We therefore examined how the system responds to temporally varying substrate supply.

In addition to constant substrate production, we imposed a pulsed input in which the substrate generation rate $\beta(t)$ periodically switches between zero and a high value, producing a square-wave modulation. This protocol mimics environments in which signaling inputs are intermittent rather than smoothly varying. Under these conditions, we tracked the time evolution of the mean substrate and product copy numbers, $\langle S \rangle$ and $\langle P \rangle$, as well as the mutual information between the internal states of enzymes $A$ and $B$, $\mathrm{MI}_{AB}$.

\begin{figure}
\centering 
\includegraphics[width=.95\linewidth]{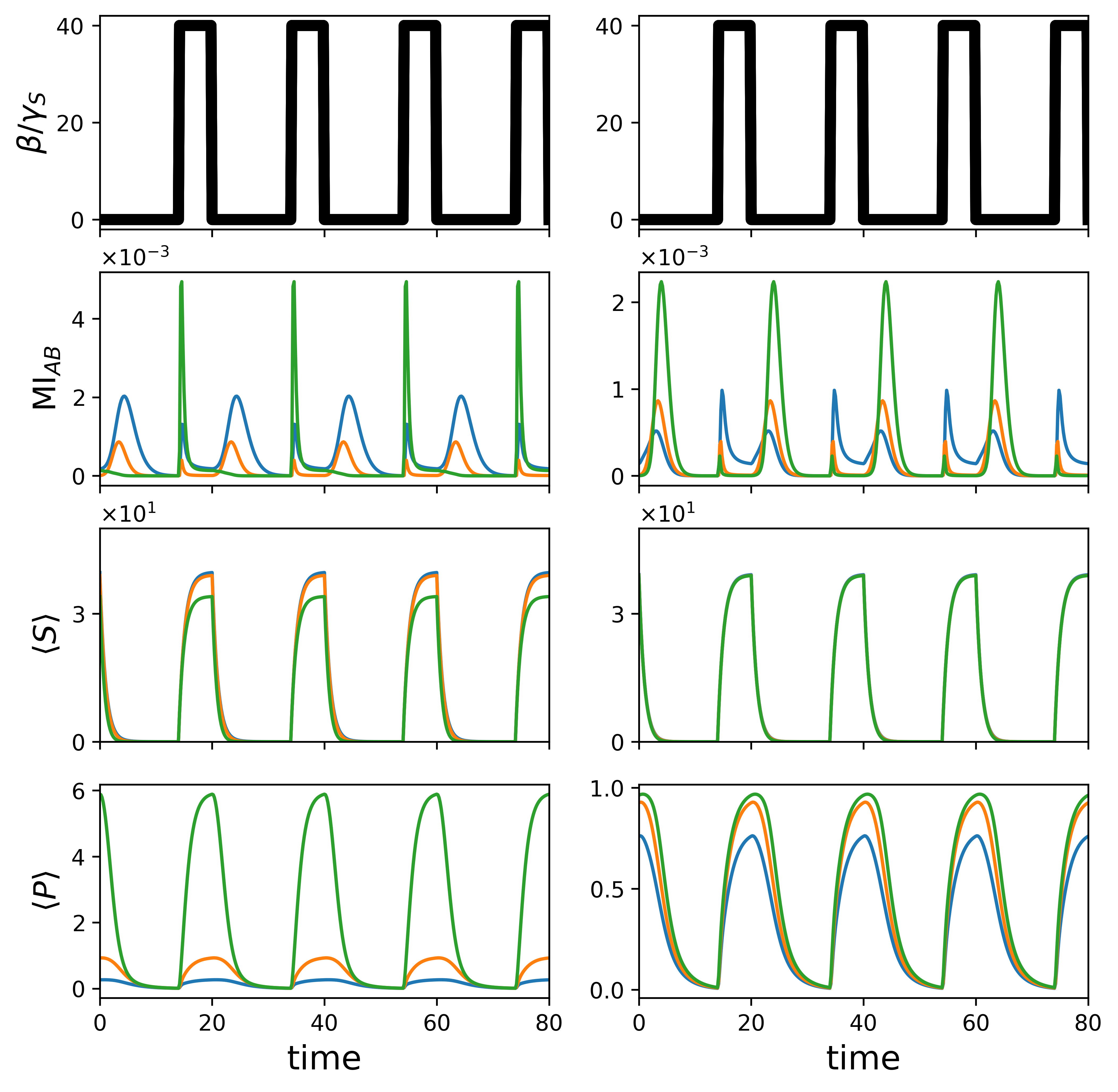} \vspace{-.5cm}
\caption{\textbf{Allostery regulates coupling between enzyme, $A$, and the downstream protein, $B$, in response to external variation in the substrate concentration, thereby providing a time order for downstream signaling.} Time traces of key metrics with varying substrate relative production rates, $\beta/\gamma_S$, change in time in cycles as shown in the top panel.
On the left-hand side, we fix K-allostery to be silent ($\xi_K = 1$) and vary V-allostery: blue denotes negative V-allostery ($\xi_V = 0.1$), orange denotes silent V-allostery ($\xi_V = 1$), and green denotes positive V-allostery ($\xi_V = 10$). On the right-hand side, we show the symmetric case, where V-allostery is silent ($\xi_V = 1$) and K-allostery is varied analogously.
The mutual information between the states of $A$ and $B$ spikes shortly after the increase in $\beta/\gamma$, even when $\langle P \rangle$ shows no significant changes. Interestingly, and consistent with the results in Fig. \ref{fig:main_steady}, the increase in mutual information is more pronounced at lower allosteric rates.} \vspace{-.6cm}\label{fig:main_variable}
\end{figure}

Fig. \ref{fig:main_variable} shows that each increase in substrate supply induces a rapid and transient spike in $\mathrm{MI}_{AB}$, reflecting a temporary strengthening of the coupling between upstream and downstream components. Notably, these information bursts often occur before substantial accumulation of product, and in some cases even when changes in $\langle P \rangle$ are minimal. This demonstrates that temporal information transfer can be dynamically regulated independently of mean product output.

The magnitude and sharpness of these information spikes depend strongly on the allosteric regime. Consistent with the steady-state behavior observed in  Figs. \ref{fig:MI_steady} and \ref{fig:main_steady}, inhibitory allostery produces particularly pronounced temporal correlations, as slower catalysis prevents immediate saturation and preserves sensitivity to upstream fluctuations. In contrast, strongly activating regimes generate faster but shorter-lived coupling events. 

Taken together, these results show that allostery regulates signaling not only by controlling reaction rates or product abundance, but by shaping when information is transmitted between pathway components. Importantly, this mechanism does not require changes in protein identity, pathway topology, or intrinsic allosteric parameters. Instead, the same biochemical architecture can be placed into distinct dynamical regimes through time-dependent modulation of substrate availability, thereby altering the temporal window over which upstream and downstream components are coupled.

This provides a physical mechanism by which cells sharing the same genome and expressing the same signaling proteins can nevertheless generate distinct functional responses. By operating identical pathways in different temporal regimes, cells can encode cell-type–specific signaling programs through timing rather than molecular specialization. In this view, differentiation and functional specificity can emerge from dynamical control of information flow, even when the underlying biochemical machinery is conserved.

\section{Conclusion}
Allostery is traditionally viewed as a mechanism for modulating enzymatic activity or product abundance. Here, we show that this view is incomplete. By combining stochastic kinetics with information theory, we demonstrate that allosteric regulation fundamentally controls when and for how long information is transmitted between signaling components. Rather than acting as a simple gain control, allostery tunes the temporal operating regime of signaling pathways, shaping the timing, duration, and coherence of fluctuations that propagate downstream.

A central implication of our results is that temporal information flow provides a powerful axis of regulation that does not require changes in molecular identity, pathway topology, or even static allosteric architecture. Cells that share the same genome, express the same signaling proteins, and operate the same biochemical pathways can nevertheless generate distinct signaling outcomes by exploiting differences in substrate availability and the resulting temporal coupling between pathway components. In this framework, differentiation does not require rewiring signaling networks, but instead emerges from dynamic control over how conserved pathways are driven in time.

This perspective offers a physical explanation for a long-standing biological puzzle: how cells with identical DNA and highly conserved signaling machinery, such as those found across tissues within an organism, can exhibit radically different behaviors and fates. Our results suggest that cell-type specificity can arise from differences in the temporal structure of signaling, even when the underlying proteins and their allosteric properties are unchanged. By selectively engaging conserved pathways in distinct dynamical regimes, cells can encode fate decisions, functional specialization, and responsiveness to environmental cues without altering the molecular components themselves.

More broadly, our work reframes allostery as a mechanism for regulating temporal information processing in noisy biochemical environments. By controlling the duration and timing of coupling between upstream and downstream proteins, allosteric regulation enables conserved signaling architectures to support diverse physiological timescales and functional outcomes. This provides a parsimonious route to biological complexity: evolution and development can diversify function not by changing what pathways are present, but by tuning how they operate in time.

\section*{Acknowledgments}
SBO acknowledges support from the Gordon and Betty Moore Foundation (AWD00034439) and National Institutes of Health (R01GM147635-01). 
SP acknowledges support from the National Institutes of Health (R35GM148237), Army Research Office (W911NF-23-1-0304), and National Science Foundation (Grant No. 2310610).

%% file: appendixs.tex
\appendix

\counterwithin{figure}{section}
\renewcommand{\thefigure}{A\arabic{figure}}
\renewcommand{\thetable}{A\arabic{table}}

\section*{Supplemental Information to ``Allostery Beyond Amplification: Temporal Regulation of Signaling Information'' }

In this Supplemental Information (SI) we explain the chemical master equation (CME) for the allosteric model used in the present study. From the set of chemical reactions in Fig.~\ref{fig:reaction_model} in the main text, we construct (in SI Sec.~\ref{SIsec:cme}) the chemical master equation (CME) that explains how the counts of each chemical species in the allosteric model change in time. Later (SI Sec.~\ref{SIsec:solution}) we show how we solved the CME in order to produce the graphs in Figs.~\ref{fig:main_steady} and \ref{fig:main_variable} and give more details of how the metrics presented there were calculated.

\section{The chemical master equation for allostery model}\label{SIsec:cme}

Chemical kinetics models are often described using average quantities derived from mass-action laws or through a purely thermodynamic framework. However, accurately capturing information transfer via allostery \emph{requires} accounting for the stochastic time evolution of the system. To achieve this, we implement a CME for the reaction network illustrated in Fig.~\ref{fig:reaction_model}. The modeled reactions, along with their stoichiometry, parameters, and kinetic rates are fully detailed in Table~\ref{tab:reactions}.

First, we enumerate all possible states of the model by assigning a unique index $i$ to each combination of the model variables $(\sigma_A, \sigma_B, P, S)$. To accomplish this, we need to set maximum counts for $S$ and $P$, which we denote by $I_S$ and $I_P$, respectively. The index $i$ for a given combination of $(\sigma_A, \sigma_B, P, S)$ is then defined as:
\begin{equation}\label{enumeration}
    i(\sigma_A, \sigma_B, P, S) =  \sigma_A (2 I_P I_S) + \sigma_B (I_P I_S) + P I_S + S.
\end{equation}

To recover the values of $\sigma_A$, $\sigma_B$, $P$, and $S$ from a given index $i$, we define the following inverse mapping functions:
\begin{subequations}\label{inverse_enumeration}
\begin{align}
    s_A(i) &= \left\lfloor \frac{i}{2 I_P I_S} \right\rfloor, \\
    s_B(i) &= \left\lfloor \frac{i}{I_P I_S} \right\rfloor \mod 2, \\
    s_P(i) &= \left\lfloor \frac{i}{I_S} \right\rfloor \mod I_P, \\
    s_S(i) &= i \mod I_S.
\end{align}
\end{subequations}
Here, $x \mod q$ denotes the remainder when $x$ is divided by $q$, and $\lfloor x \rfloor$ represents the greatest integer less than or equal to $x$. 
We use the notation $s_A$, $s_B$, $s_P$, and $s_S$ to indicate that these are functions mapping the index $i$ back to the model variables, distinguishing them from the raw values $\sigma_A$, $\sigma_B$, $P$, and $S$. 
In particular, when referencing a state by its index $i$, this indexing scheme ensures that each function $s_A(i)$, $s_B(i)$, $s_P(i)$, and $s_S(i)$ will return the corresponding value of the model variables ($\sigma_A$, $\sigma_B$, $P$, and $S$) for that indexed state. That is, if a state is indexed by $i$, then $s_A(i) = \sigma_A$, $s_B(i) = \sigma_B$, $s_P(i) = P$, and $s_S(i) = S$.

With this indexing framework in place, we construct the CME with two elements. The first is the probability vector, which we represent as a row vector $\bar{\rho} = (\rho_1,\rho_2, \ldots, \rho_I)$ where $\rho_i$ is the probability of the system being in state $i$ and $I$ is the total number of modeled states, $I = 8 I_S I_P$. As we want to observe the system's time evolution we write $\bar{\rho}(t)$. 

The second is the propagator matrix, whose construction is as follows. For each state, indexed by $i$, we identify all potential resultant state indices ($j$), which are determined by changes in state indicated in the central columns of Table \ref{tab:reactions}. 
The rate matrix $\mathbf{\Lambda}$, composed of elements  $\lambda_{ij}$, is then defined such that $\lambda_{ij}$ represents the transition rate (specified in the rightmost column of Table \ref{tab:reactions}).
The complete rate matrix depends on all the kinetic parameters of the model, which we will collectively denote by $\theta$, making each element a function of these parameters: $\lambda_{ij}(\theta)$. To simplify notation, we write $\mathbf{\Lambda}_\theta$ for the matrix. 
This rate matrix is then used to construct the propagator matrix $\mathbf{G}_\theta$ with elements $g_{ij}(\theta)$ defined as 
\begin{equation}
        g_{ij}(\theta) \equiv
    \begin{cases}
    -\sum_k  \lambda_{ik} \quad & \text{if} \ i=j \\ 
    \lambda_{ij} \quad & \text{otherwise}
    \end{cases} \ .
\end{equation}
It is important to note that for each line (or state) $i$, there are only up to 15 possible reactions (corresponding to the reactions listed in Table~\ref{tab:reactions}), thus only up to 15 non-zero $\lambda_{ij}$ out of $I$ entries in that row, making $\mathbf{G}_\theta$ a sparse matrix. 

The time evolution of the states' probability is given by the differential equation~\cite{VanKampen,Presse23,Pessoa24}
\begin{equation}\label{cme}
    \dv{\bar{\rho}}{t} = \bar{\rho} \mathbf{G}_\theta(t) \ .
\end{equation}
In the following subsection we show how we set up the kinetic parameters $\theta$, the initial condition $\bar{\rho}$, and how to solve \eqref{cme} by employing methods that leverage the fact that we are dealing with a sparse $\mathbf{G}_\theta$~\cite{Pessoa24}. 
In short, to construct Fig.~\ref{fig:main_steady} we evolve \eqref{cme} towards its steady-state ($\bar{\rho} \mathbf{G}_\theta=0$) while to generate Fig.~\ref{fig:main_variable} we evolve \eqref{cme} with a varying production rate of $S$, $\beta$. The details are outlined in the following section.

\begin{table}
\caption{Table summarizing the chemical reactions within the minimal allosteric model, as illustrated in Fig.~\ref{fig:reaction_model}. Each line represents one of the reactions, its parameters, and the transition rates between the states before and after the reaction, with $\delta$ here representing the Kronecker delta. The central columns represent the change in each of the model variables --- labeled as $\sigma_A$, $\sigma_B$, $P$, and $S$ --- and how they are changed when the respective reaction happens. These values are required to write the rate matrix, as described in SI Sec.~\ref{SIsec:cme}. }
\label{tab:reactions}
\centering
\begin{tabular}{rcl||c|c|c|c||cl}
\multicolumn{3}{c||}{Reaction}     & $\sigma_A$ & $\sigma_B$ & $P$  & $S$  & Parameters             & Rates                                   \\ 
\hline
$\emptyset$ & $\to$ & $S$          &            &            &      & $+1$ & $\beta$                & $\beta$                                       \\
$S$         & $\to$ & $\emptyset$  &            &            &      & $-1$ & $\gamma_S$             & $\gamma_S S$                                  \\
$A + S$     & $\to$ & $AS$         & $0 \to 2$  &            &      & $-1$ & $k_{A \text{on}}$      & $k_{A \text{on}} S \ \delta_{\sigma_A}^{0}$   \\
$AS$        & $\to$ & $A + S$      & $2 \to 0$  &            &      & $+1$ & $k_{A \text{off}}$     & $k_{A \text{off}} \ \delta_{\sigma_A}^{2}$    \\
$A^\ast +S$ & $\to$ & $A^\ast S$   & $1 \to 3$  &            &      & $-1$ & $k_{A^\ast \text{on}}$ & $k_{A^\ast \text{on}} S \  \delta_{\sigma_A}^{1}$ \\
$A^\ast S$  & $\to$ & $A^\ast + S$ & $3 \to 1$  &            &      & $+1$ & $k_{A^\ast \text{off}}$& $k_{A^\ast \text{off}}\  \delta_{\sigma_A}^{3}$ \\
$A$         & $\to$ & $A^\ast$     & $0 \to 1$  &            &      &      & $\alpha$               & $\alpha \ \delta_{\sigma_A}^{0}$                 \\
$A^\ast$    & $\to$ & $A$          & $1 \to 0$  &            &      &      & $\alpha^\ast$          & $\alpha^\ast \ \delta_{\sigma_A}^{1}$            \\
$AS$        & $\to$ & $A^\ast S$   & $2 \to 3$  &            &      &      & $\alpha_S$             & $\alpha_S \ \delta_{\sigma_A}^{2}$               \\
$A^\ast S$  & $\to$ & $AS$         & $3 \to 2$  &            &      &      & $\alpha^\ast_S$        & $\alpha^\ast_S \ \delta_{\sigma_A}^{3}$          \\
$AS$        & $\to$ & $A$          & $2\to 0$   &            & $+1$ &      & $\nu$                  & $\nu \ \delta_{\sigma_A}^{2}$                    \\
$A^\ast S$  & $\to$ & $A^\ast$     & $3 \to 1$  &            & $+1$ &      & $\nu^\ast$             & $\nu^\ast  \ \delta_{\sigma_A}^{3}$              \\
$B+P$       & $\to$ & $BP$         &            & $0 \to 1$  & $-1$ &      & $k_{B \text{on}}$      & $k_{B \text{on}} P \ \delta_{\sigma_B}^{0}$     \\
$BP$        & $\to$ & $B+P$        &            & $1 \to 0$  & $+1$ &      & $k_{B \text{off}}$     & $k_{B \text{off}} \ \delta_{\sigma_B}^{1}$      \\
$P$         & $\to$ & $\emptyset$  &            &            & $-1$ &      & $\gamma_P$             & $\gamma_P P$     
\\ 
\hline                            
\end{tabular}
\end{table}

\section{Solving the chemical master equation}\label{SIsec:solution}
To clearly explain the methodologies behind the generation of the main text figures, and consequently the conclusions drawn in our study, this section details the process for solving the CME on the minimal allosteric model. 
In order to do so we need to define the initial condition, $\bar{\rho}$, and the set of parameters, $\theta$, needed to construct the propagator matrix $\mathbf{G}_\theta$ and solve \eqref{cme} for that matrix. However, in the main text we had two different goals, and these translate into two different types of information we want to draw from the CME solution. 

The first goal is to find the steady state, $\dv{\bar{\rho}}{t} = 0$, for a constant generation rate of $S$, denoted as $\beta$, and varying values for the allosteric rate $\xi = \frac{\nu^\ast}{\nu}$. This generates Figs.~\ref{fig:MI_steady} and ~\ref{fig:main_steady} in the main text. We explain the initial condition and the process to find the steady state in SI Sec.~\ref{SIsec:const_beta}.

The second goal, which generates Fig.~\ref{fig:main_variable} is to solve the CME for a production rate changing in time. We explain the initial condition and how to solve \eqref{cme} and, consequently, obtain the time evolution of the metrics in SI Sec.~\ref{SIsec:variable_beta}.

In both cases the degradation rate of $S$ is set as $\gamma_S=1$, or equivalently $1/\gamma_S$ is the unit of time. In both cases we also show the effect of varying the allosteric rate $\xi$. 
We also look at how changes in the $\xi$ rate affect our results. We list all the parameters we used in the main text. In both cases we also explore an equivalent non-allosteric system designed to closely replicate the allosteric kinetics. For it we keep the same model but guarantee that the allosteric states ($A^\ast$ and, consequently, $A^\ast S$) are never reached by the system by fixing the transition to those states as zero.Finally, we explain the metrics used to visualize the steady-state solutions of the first goal and the time evolution of the second goal based on the state probabilities obtained from \eqref{cme} in SI Sec.~\ref{SIsec:metrics}.

\subsection{First goal --- steady state of constant $\beta$}\label{SIsec:const_beta}
To construct the initial condition we look into the steady-state of $S$ if the allosteric enzyme were not present. If we restrict the reactions in Table \ref{tab:reactions} to the first two lines we obtain a birth–death process with respective rates $\beta$ and $\gamma_S$, thus the steady state would be a Poisson distribution for $S$ with rate $\frac{\beta}{\gamma_S}$. 
In the cases where $\beta$ is constant (Fig.~\ref{fig:main_steady}) we use an initial condition of $\sigma_A$ in the non-allosteric state $A$ ($\sigma_A = 0$), $\sigma_B$ in the state of unattached $B$ ($\sigma_B=0$), initially no product in the system ($P=0$), and a Poisson distribution with rate $\frac{\beta}{\gamma_S}$ for $S$. Note that this is done for computational convenience and does not change the steady state.

We evolve the CME using the method described as R-MJP in \cite{Pessoa24}, by using the probabilities described in the previous paragraph as the initial condition $\bar{\rho}(0)$ and then calculating the probability vector at different times $\bar{\rho}(t)$ in intervals of $10$, $t = \{10,20,30, \ldots \}$. We claim that we have found the steady state when we find a time such that the absolute value of the maximum element in the left-hand side of \eqref{cme}, $\bar{\rho}(t) \mathbf{G}_\theta$, is smaller than $10^{-12}$.



\subsection{Second goal --- time evolution for variable $\beta$}\label{SIsec:variable_beta}

In the main text, when studying a production rate $\beta$ that varies in time (Figs.~\ref{fig:main_variable}, we consider a pulsed production in which the substrate production rate periodically switches between zero and a high value.

In this ``on--off'' scheme, $\beta(t)$ takes the maximal value $\beta_{\max}$ for a finite portion of each period $\tau$, and remains zero for the remainder:
\begin{equation}
    \beta(t) =
    \begin{cases}
        \beta_{\max} \quad & \text{if} \ \ t \mod \tau > t_i, \\
        0 \quad & \text{otherwise} ,
    \end{cases}
\end{equation}
where $t_i$ denotes the duration of the ``off'' interval within each period. 
Thus, within every cycle of length $\tau$, $\beta(t)$ alternates between $0$ and $\beta_{\max}$, generating a square-wave modulation in substrate supply. 
An example of this pulsed protocol is shown in the bottom panel of Fig.~\ref{fig:main_variable}, where we use $\beta_{\max} = 40$, $\tau = 20$, and $t_i = 16$.

To implement this time dependence in the CME, we construct a time-dependent generator $\mathbf{G}_\theta(t)$ using the above definition of $\beta(t)$ while keeping all other kinetic parameters fixed. 
The initial condition is set as described in the previous subsection, with simulations starting at time $t = -10$ to minimize the influence of initialization transients. 
In Fig.~\ref{fig:main_variable} we present results over four full periods, $4\tau$.

Because $\beta(t)$ is piecewise constant within each on/off segment, we divide the timeline at the discontinuity points and treat each interval independently, allowing us to apply the same stationary-propagator method used for constant $\beta$, referred to as R-MJP in \cite{Pessoa24}.

\subsection{Metrics}\label{SIsec:metrics}
Representing the solution of the master equation means obtaining the probabilities of every possible state. 
In Figs.~\ref{fig:MI_steady}--\ref{fig:main_variable} we present the solution in terms of three metrics. 

The first metric we use is the mutual information between the states of $A$ and $B$, defined in \eqref{mi_def}. To compute this, we need the marginalized joint probability distribution $p(\sigma_A, \sigma_B)$, which represents the probability of $A$ being in state $\sigma_A$ and $B$ being in state $\sigma_B$.
Note that, so far, we have discussed how to calculate the probabilities of the overall states, enumerated by the index $i$, in the previous subsections (\ref{SIsec:const_beta} and \ref{SIsec:variable_beta}). For each state indexed by $i$, we can determine the states of $A$ and $B$ using the functions $s_A(i)$ and $s_B(i)$, as defined in \eqref{inverse_enumeration}. Thus, we can express the joint probability distribution $p(\sigma_A, \sigma_B)$ as a sum over all states $i$:
\begin{equation}
    p(\sigma_A, \sigma_B) = \sum_i \rho_i \, \delta_{\sigma_A}^{s_A(i)} \delta_{\sigma_B}^{s_B(i)} \ ,
\end{equation}
since $\rho_i$ is the probability of being in the state indexed by $i$, and $\delta_{\sigma_A}^{s_A(i)}$ and $\delta_{\sigma_B}^{s_B(i)}$ ensure that only states with $s_A(i) = \sigma_A$ and $s_B(i) = \sigma_B$ contribute to the sum.
Once we have the joint distribution $p(\sigma_A, \sigma_B)$, we can calculate the marginal distributions $p(\sigma_A)$ and $p(\sigma_B)$ by summing over the other variable:
\begin{align}
    p(\sigma_A) &= \sum_{\sigma_B} p(\sigma_A, \sigma_B) = \sum_i \rho_i \, \delta_{\sigma_A}^{s_A(i)}  \ , \\
    p(\sigma_B) &= \sum_{\sigma_A} p(\sigma_A, \sigma_B) = \sum_i \rho_i \,  \delta_{\sigma_B}^{s_B(i)} \ .
\end{align}
allowing us to then calculate $\text{MI}_{AB}$ from \eqref{mi_def} as
\begin{equation}\label{mi_def}
\text{MI}_{AB} = \sum_{\sigma_A} \sum_{\sigma_B}  \left(\sum_i \rho_i \, \delta_{\sigma_A}^{s_A(i)} \delta_{\sigma_B}^{s_B(i)}\right) \left(  \log \left( \sum_i \rho_i \, \delta_{\sigma_A}^{s_A(i)} \delta_{\sigma_B}^{s_B(i)} \right) - \log\left(\sum_j \rho_j \delta_{\sigma_A}^{s_A(j)}  \right) - \log\left(\sum_k \rho_k \delta_{\sigma_B}^{s_B(k)}\right) \right)  \, .
\end{equation}  

The other two metrics we are interested in are the expected values of $S$ and $P$, which represent the average values of these variables over all states. Similarly to how we calculated $p(\sigma_A, \sigma_B)$, we can obtain the marginal distributions for the variables $S$ and $P$ as:
\begin{equation}
    p(S) = \sum_i \rho_i \, \delta_{S}^{s_S(i)} \ , \quad \text{and} \quad p(P) = \sum_i \rho_i \, \delta_{P}^{s_P(i)} \ .
\end{equation}
Once we have these marginal distributions $p(S)$ and $p(P)$, we can calculate the expected values $\langle S \rangle$ and $\langle P \rangle$, defined as:
\begin{equation}
    \langle S \rangle = \sum_{S} p(S) S  = \sum_i  \rho_i s_S(i)\quad \text{and} \quad \langle P \rangle = \sum_{P}  p(P) P = \sum_i \rho_i s_P(i) \ .
\end{equation}
In summary, from the probability distribution $\rho_i$ over states indexed by $i$ --- obtained from solving the CME --- we can derive all three metrics: the mutual information $\text{MI}_{AB}$, and the expected values $\langle S \rangle$ and $\langle P \rangle$.